\documentclass[nonacm,sigconf]{acmart}
\usepackage{ascmac}
\usepackage{booktabs}
\usepackage{subfigure}
\usepackage[norelsize,ruled,vlined,linesnumbered]{algorithm2e}
\usepackage{algorithmic}

\newtheoremstyle{mystyle} 
    {1.5mm}
    {1.5mm}
    {\it}
    {0mm}
    {\scshape}
    {.}
    { }
    {}
\theoremstyle{mystyle}

\newtheorem{definition}{Definition}
\newtheorem{example}{Example}
\newtheorem{lemma}{Lemma}
\newtheorem{theorem}{Theorem}

\newcommand{\wsq}{\hspace{\fill}$\square$}
\newcommand{\vsp}{\vspace{1.5mm}}
\SetKwComment{Comment}{$\triangleright$\ }{}

\AtBeginDocument{%
  }

\setcopyright{rightsretained}
\copyrightyear{2023}
\acmYear{2023}
\acmDOI{10.1145/3603719.3603720}
\acmISBN{979-8-4007-0746-9/23/07}
\acmConference[]{}{}{}
\acmBooktitle{}

\makeatletter
\gdef\@copyrightpermission{
  \begin{minipage}{0.3\columnwidth}
   \href{https://creativecommons.org/licenses/by-nc/4.0/}{\includegraphics[width=0.90\textwidth]{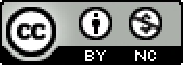}}
  \end{minipage}\hfill
  \begin{minipage}{0.7\columnwidth}
   \href{https://creativecommons.org/licenses/by-nc/4.0/}{This work is licensed under a Creative Commons Attribution-NonCommercial International 4.0 License.}
  \end{minipage}
  \vspace{5pt}
}
\makeatother

\setcopyright{none}
\renewcommand\footnotetextcopyrightpermission[1]{} 
\begin{document}

\title{Fast Algorithm for Embedded Order Dependency Validation (Extended Version)}

\author{Alejandro Ramos}
\affiliation{%
  \institution{Osaka University}
  \country{Japan}
}
\email{ramos.alejandro@ist.osaka-u.ac.jp}

\author{Takuya Uemura}
\affiliation{%
  \institution{Osaka University}
  \country{Japan}
}
\email{uemura.takuya@ist.osaka-u.ac.jp}

\author{Daichi Amagata}
\affiliation{%
  \institution{Osaka University}
  \country{Japan}
}
\email{amagata.daichi@ist.osaka-u.ac.jp}

\author{Ryo Shirai}
\affiliation{%
  \institution{Osaka University}
  \country{Japan}
}
\email{shirai.r@ist.osaka-u.ac.jp}

\author{Takahiro Hara}
\affiliation{%
  \institution{Osaka University}
  \country{Japan}
}
\email{hara@ist.osaka-u.ac.jp}

\begin{abstract}
Order Dependencies (ODs) have many applications, such as query optimization, data integration, and data cleaning.
Although many works addressed the problem of discovering OD (and its variants), they do not consider datasets with missing values, a standard observation in real-world datasets.
This paper introduces the novel notion of Embedded ODs (eODs) to deal with missing values.
The intuition of eODs is to confirm ODs only on tuples with no missing values on a given embedding (a set of attributes).
In this paper, we address the problem of validating a given eOD.
If the eOD holds, we return true.
Otherwise, we search for an updated embedding such that the updated eOD holds.
If such embedding does not exist, we return false.
A trivial requirement is to consider an embedding such that the number of ignored tuples is minimized.
We show that it is NP-complete to compute such embedding.
We therefore propose an efficient heuristic algorithm for validating embedded ODs.
We conduct experiments on real-world datasets, and the results confirm the efficiency of our algorithm.
\end{abstract}

\maketitle

\section{Introduction}  \label{sec:introduction}

\subsection{Motivation}
Integrity constraints are commonly used in a number of applications in data profiling \cite{14}, such as data integration and cleaning \cite{Fan2008Conditional}, query optimization \cite{szlichta2012Expressiveness}, and schema design \cite{16}.
Among integrity constraints in relational databases, the most important constraints are functional dependencies (FDs) \cite{15} and order dependencies (ODs) \cite{ginsburg1983order,szlichta2012fundamentals}.
Informally, FDs describe that the values of some attributes functionally determine the value of others.
An FD $X\to Y$ says that, for each tuple in a database, the values of the attributes in $X$ determine the values of the attributes in $Y$.
ODs describe that the \textit{order} of tuples w.r.t. given attributes determines the order of tuples w.r.t. other attributes.
An OD $X \mapsto Y$ says that, if all values in $X$ are increasing (or decreasing), then all values in $Y$ must also be increasing (or decreasing).
Because FDs are a special case of ODs \cite{szlichta2012Expressiveness}, this paper focuses on ODs.

Many works addressed the problem of discovering ODs \cite{jin2022efficient,langer2016efficient,szlichta2016effective}, because ODs help query optimization, violation detection, and data repairing, to name a few.
It is well-known that real-world datasets contain errors \cite{amagata2021dpc,amagata2021fast,amagata2022fast,amagata2023diversity,amagata2023efficient}. 
To deal with erroneous data, there are some variants of ODs, such as approximate ODs \cite{jinA2021aproximate}.
Surprisingly, although missing values are standard observations in real-world data, no existing works have considered ODs on data with missing values.
We therefore consider how to define ODs when datasets contain missing values, and use the concept \textit{Embedded FDs} \cite{wei2021algorithms}.
This concept defines FDs by extracting only complete data on some embedded attributes (i.e., tuples that have no missing values on the attributes). 
This idea helps find useful ODs that are valid semantically in relations, whereas they may be invalid in the other OD definitions.
The found dependencies can be used to define integrity requirements and completeness requirements.

We consider, in Table \ref{tab:employee}, OD Rank $\mapsto$ Salary; Salary increases as Rank increases.
This is a \textit{semantically} valid OD, but cannot be found by the existing OD definition because of the missing value (denoted by $\perp$) of $t_2$. 
By considering a ``sub-table'', where only no missing values are permitted in Salary, i.e., $t_2$ is removed from the relation, Rank $\mapsto$ Salary holds.
This example demonstrates the effectiveness of embedded ODs (eODs), and they can find meaningful ODs that cannot be found from the original OD definition.
Therefore, this paper considers an algorithm that, given a pair of attribute lists, returns its OD validity and an embedding which makes the pair valid under the embedding (if such an embedding exists).
The formal problem definition appears in Section \ref{sec:preliminary}.

\begin{table}[t]
  \centering
  \caption{Employee list}
  \label{tab:employee}
  \begin{tabular}{ccccc} \toprule
      ID    & Rank  & Years & Age   & Salary    \\ \midrule
      $t_1$ & 1     & 1     & 20    & 15000     \\ 
      $t_2$ & 2     & 1     & 21    & $\perp$   \\ 
      $t_3$ & 3     & 2     & 22    & 25000     \\ 
      $t_4$ & 4     & 3     & 25    & 30000     \\ \bottomrule
  \end{tabular}
\end{table}

\subsection{Challenge}
Actually, checking whether a given OD candidate (a pair of attributes) is valid is not a difficult task, because a state-of-the-art algorithm OD \cite{langer2016efficient} achieves this.
However, \textit{efficiently computing an embedding on which the pair of attributes is valid is not trivial}.
A straightforward approach is to enumerate all possible embeddings (i.e., attribute lists) and check the validity under the embeddings by using the state-of-the-art algorithm, see Section \ref{sec:preliminary}.
Trivially, this approach incurs a \textit{factorial} cost w.r.t. the number of attributes, so this does not scale well to large relational tables.

As noted above, ODs are used in, for example, query optimization.
If embedding computation incurs a substantial computational cost, we cannot support query optimization, as embedding computation can be the main bottleneck.
Therefore, an algorithm that efficiently solves our problem is required.

\subsection{Contribution}
This paper proposes an efficient algorithm that overcomes the above challenge.
To compute a valid embedding, we can focus only on violation tuples and their attributes with missing values.
Our task then becomes an evaluation of whether these tuples can be removed by adding these attributes to an embedding.
This approach can reduce the search space and avoid the factorial cost.

To summarize, this paper makes the following contributions.
We
\begin{itemize}
    \setlength{\leftskip}{-4.0mm}
    \item   formulate the novel concept of embedded order dependencies,
    \item   present an algorithm that efficiently checks whether or not an OD holds in some embedding, and
    \item   evaluate the performance of our proposed algorithm on real-world datasets.
\end{itemize}

\vsp
\noindent
\underline{\textbf{Comparison with our conference version.}}
The above contents appear in our conference version \cite{amagata2023fast}.
When considering embedding (i.e., a sub-table that contains no missing values), applications would be happy if the sub-table size is maximized, i.e., the number of ignored tuples is minimized.
Then, a natural requirement is to consider an embedding that yields the sub-table.
However, as we show in this paper, it is NP-complete to find the sub-table.
The proof of this hardness is a new contribution.
Providing this fact further justifies the design of our heuristic algorithm.

\section{Preliminaries} \label{sec:preliminary}

\subsection{Problem Definition}
We use $\mathbf{R}$ to denote a relational schema, and $\mathbf{r}$ is a specific relational instance (table).
Also, we use $A$ ($B$) to denote an attribute of $\mathbf{r}$, whereas $\mathbf{X},\mathbf{Y}$ are \emph{lists} of attributes.
Let $s,t$ represent tuples $\in \mathbf{r}$, and $s_A$ is the value of $s$ on $A$.
For ease of understanding, we first define order dependencies.

\begin{definition}[\textsc{Order dependencies} \cite{szlichta2012fundamentals}]
An order dependency over a schema $\mathbf{R}$ is a statement of the form $\mathbf{X}\mapsto_{\leq}\mathbf{Y}$, where $\mathbf{X},\mathbf{Y} \in \mathbf{R}$.
An order dependency $\mathbf{X}\mapsto_{\leq}\mathbf{Y}$ is valid iff for any two tuples $s,t \in \mathbf{r}, s_X \leq t_X \Rightarrow s_Y \leq t_Y$\footnote{Actually, extending to $s_X \geq t_X \Rightarrow s_Y \leq t_Y$, $s_X \leq t_X \Rightarrow s_Y \geq t_Y$, and $s_X \geq t_X \Rightarrow s_Y \geq t_Y$ is also possible.
For ease of presentation, we use the increasing order.}.
\end{definition}

Next, let $\mathbf{E}$ be a subset of attributes of $\mathbf{R}$.
We call $\mathbf{E}$ \emph{embedding}, and it defines $\mathbf{r}^\mathbf{E} \in \mathbf{r}$, which is a set of tuples with no missing values on $\mathbf{E}$.
Then, we define a new concept, \textit{embedded order dependencies}.

\begin{definition}[\textsc{Embedded order dependencies}]
An embedded order dependency (eOD) over a relational schema $\mathbf{R}$ is a statement of the form $\mathbf{E}: \mathbf{X}\mapsto_{\leq}\mathbf{Y}$, where $\mathbf{X},\mathbf{Y} \subseteq \mathbf{E} \subseteq \mathbf{R}$.
An eOD  $\mathbf{E}: \mathbf{X}\mapsto_{\leq}\mathbf{Y}$ is valid iff for any two tuples $s,t \in \mathbf{r}^{\mathbf{E}}, s_X \leq t_X \Rightarrow s_Y \leq t_Y$.
\end{definition}

\noindent
This paper considers the following problem:
\begin{description}
    \setlength{\leftskip}{-3.0mm}
    \item[\textbf{Input}]   Given a statement $\mathbf{E}: \mathbf{X} \mapsto_{\leq} \mathbf{Y}$, check whether it is valid.
    \item[\textbf{Output}]  If true, return \textbf{valid}.
                            Otherwise, one of the following options is returned:
                            (1) \textbf{valid with} $\mathbf{E}'$ iff there is an embedding $\mathbf{E}' \supset \mathbf{E}$ such that $\mathbf{E}': \mathbf{X} \mapsto_{\leq} \mathbf{Y}$ holds, and
                            (2) \textbf{not valid} if there does not exist such $\mathbf{E}'$.
\end{description}

\begin{example} 
If $AB: A \mapsto_{\leq} B$ holds, this problem returns ``valid.''
If it does not, this problem searches for an embedding $\mathbf{E}' \supset AB$ such that $\mathbf{E}': A \mapsto_{\leq} B$ holds.
Then, if $ABC: A \mapsto_{\leq} B$ holds, this problem returns ``valid with $ABC$.''
On the other hand, if there is no  $\mathbf{E}' \supset AB$ such that $\mathbf{E}': A \mapsto_{\leq} B$ holds, this problem returns ``not valid.''
\end{example}

Before solving the above problem, we introduce two important concepts \emph{split} and \emph{swap} \cite{szlichta2012fundamentals}.
They are used as remarks that an OD under ``$\leq$'' can be invalid.
In addition, in \cite{langer2016efficient}, the concept of \emph{merge} is introduced, and it makes an OD under ``$<$'' invalid.
We formally define them below.

\begin{definition}[\textsc{Split}]
Given tuples $s,t \in \mathbf{r}$ and attributes $A,B \in \mathbf{R}$, there is a split if $s_A=t_A$ but $s_B \neq t_B$.
\end{definition}

\begin{definition}[\textsc{Merge}]
Given tuples $s,t \in \mathbf{r}$ and attributes $A,B \in \mathbf{R}$, there is a merge if $s_A \neq t_A$ but $s_B = t_B$.
\end{definition}

\begin{definition}[\textsc{Swap}]
Given tuples $s,t \in \mathbf{r}$ and attributes $A,B \in \mathbf{R}$, there is a swap if $s_A < t_A$ but $s_B > t_B$.
\end{definition}

From the definition of split and merge, a split on ($A$,$B$) implies a merge on ($B$,$A$) and vice-versa.
Splits invalidate ODs under ``$\leq$'', and merges invalidate ODs under ``$<$''.
Swaps invalidate ODs under both ``$\leq$'' and ``$<$''.
Based on this observation, we introduce the validity of eODs under ``$\leq$'' and  ``$<$''.
Specifically, we have the following three lemmas according to \cite{langer2016efficient}.

\begin{lemma}
$\mathbf{E}: A \mapsto_{\leq} B$ is valid, iff there is neither a split nor a swap on attributes $A,B$ in $\mathbf{r}^{\mathbf{E}}$.
\end{lemma}

\begin{lemma}
$\mathbf{E}: A \mapsto_{<} B$ is valid, iff there is neither a merge nor a swap on attributes $A,B$ in $\mathbf{r}^{\mathbf{E}}$.
\end{lemma}

\begin{lemma}
$\mathbf{E}: A \mapsto_{<} B$ is invalid, if there is a merge or a swap on $A,B$ in $\mathbf{r}^{\mathbf{E}}$.
\end{lemma} 

\noindent
Although the above statements have the embedding $\mathbf{E}$, the validity focuses on $\mathbf{r}^{\mathbf{E}}$ and attributes $A$ and $B$.
Therefore, we can directly use these lemmas.
Then, as shown in \cite{langer2016efficient}, we have:

\begin{theorem} \label{theorem:interchange}
$\mathbf{E}: A \mapsto_{<} B$ is valid $\Leftrightarrow \mathbf{E}: B \mapsto_{\leq} A$ is valid.
\end{theorem}

\noindent
This theorem means that we only need to check for eODs under ``$<$'' for validity.
This is because if the OD under ``$<$'' is valid, it means that the OD under ``$\leq$'' is also valid (except for the case where the left-hand side, LHS, and right-hand side, RHS, are swapped).

\subsection{Na\"ive Algorithm}
We here consider a na\"ive algorithm that solves our problem.
Recall Theorem \ref{theorem:interchange}, and validating $\mathbf{E}: B \mapsto_{<} A$ is sufficient to validate $\mathbf{E}: A \mapsto_{\leq} B$.
We do this because it is easier to find errors for an eOD under ``$<$'' \cite{langer2016efficient}, and we can focus on finding where swaps and merges occur.
The na\"ive algorithm has the following steps.
\begin{enumerate}
    \setlength{\leftskip}{-3.0mm}
    \item   It finds swaps and merges for $B \mapsto_{<} A$ through \textsc{FindErrors}\footnote{This is based on the validation algorithm from ORDER \cite{langer2016efficient}.
            We extend the algorithm so that we can obtain $\mathbf{S}$ (a set of swaps) and $\mathbf{M}$ (a set of merges).
            We still use \emph{sorted partitions} \cite{langer2016efficient} to efficiently find swaps and merges.
            While the original implementation terminates whenever it finds a swap or merge, our implementation adds the tuples with this swap (merge) to $\mathbf{S}$ ($\mathbf{M}$) and continues to scan the corresponding sorted partitions.}.
    \item   It next checks whether all of these errors disappear under the given embedding through \textsc{CheckForErrorDeletion}\footnote{For each swap $s \in \mathbf{S}$ or merge $m \in \mathbf{M}$, \textsc{CheckForErrorDeletion} checks whether there is a missing value $\perp$ in an attribute in $\mathbf{E}$ on the tuple that causes the error.
            If the tuple has a missing value on the attribute, then $\mathbf{r}^{\mathbf{E}}$ does not have the tuple.
            Thus, \textsc{CheckForErrorDeletion} removes the error (tuple).}.
            If so, it returns valid.
    \item   Otherwise, for every possible embedding $\mathbf{E}' \in \mathbf{R}$ such that $\mathbf{E} \subset \mathbf{E}'$, it repeats steps 1 and 2.
            If there exists $\mathbf{E}'$ such that $B \mapsto A$ holds, it returns $\mathbf{E}'$.
    \item   If there does not exist such an $\mathbf{E}'$, it returns not valid.
\end{enumerate}

This algorithm exactly solves our problem.
However, it requires a \textit{factorial} number of checks in the worst case if the OD is invalid under the given $\mathbf{E}$.

\section{Hardness of Computing Embedding for Minimizing Ignored Tuples} \label{sec:hardness}
Although our problem does not have a constraint on $\mathbf{E}'$, it is natural to consider that the number of ignored tuples on $\mathbf{E}'$ (i.e., $|\mathbf{r}^{\mathbf{E}} - \mathbf{r}^{\mathbf{E'}}|$) should be minimized.
Unfortunately, it is hard to compute such $\mathbf{E}'$ efficiently.

\begin{theorem}
Given $\mathbf{E}: \mathbf{X} \mapsto_{\leq} \mathbf{Y}$, assume that there exits $\mathbf{E}'$ such that $\mathbf{E}': \mathbf{X} \mapsto_{\leq} \mathbf{Y}$ holds.
Then, it is NP-complete to find $\mathbf{E}'$ such that the number of ignored tuples on $\mathbf{E}'$ is minimized among all embeddings that provide $\mathbf{E}'': \mathbf{X} \mapsto_{\leq} \mathbf{Y}$.
\end{theorem}

\noindent
\textsc{Proof.}
To prove this theorem, we show that there exists an instance of this problem which can be reduced to the weighted minimum set cover problem, which is NP-complete.
The input of the weighted set cover problem is a collection of subsets $Q_{i} \subseteq P$ each of which has a positive weight $w_i$, and $P = \bigcup Q_{i}$.
The output of this problem is the collection of $Q_{j}$ such that $P = \bigcup Q_{j}$ and the sum of the weights is minimized.

Now let $w_i$ be the number of ignored tuples when $A_i$ is added into $\mathbf{E}$, i.e., $w_i = |\mathbf{r}^{\mathbf{E}} - \mathbf{r}^{\mathbf{E'}}|$.
Similarly, let $Q_j$ be a set of tuples with missing values on $A_i$.
At a first look, this setting is the same as the weighted set cover problem, but it is different.
It is important to notice that $w_i$ is generally variable and depends on the current $\mathbf{E'}$\footnote{$\mathbf{E'}$ is initialized by $\mathbf{E}$.
Whenever $\mathbf{E'} \gets \mathbf{E'} + A_i$, the tuples with missing value on $A_i$ is ignored.
Notice that such tuples may have missing values on $A_j$.}.
We hence consider the following instance:
\begin{itemize}
    \setlength{\leftskip}{-4.0mm}
    \item   $s^{i}_{X} = i$ and $s^{i}_{Y} = \lfloor \frac{i + 1}{2} \rfloor$, where $s^{i}$ represents the $i$-th tuple in a given table.
    \item   Each tuple has at most one missing value.
\end{itemize}
Then, each pair of $(s^{2j-1},s^{2j})$ causes a merge, where $j$ is an integer, and we do not have other swaps and merges.
Let $O$ be a set of integers not larger than $|\mathbf{r}|$, where $|\mathbf{r}|$ is the number of tuples in $\mathbf{r}$.
Furthermore, let $Q_i$ be a set of non-negative odd integers $j$ such that $s^j$ has a merge and is ignored if $A_{i}$ is added into $\mathbf{E}$.
In this condition, $w_i$ is not variable anymore.
Therefore, we see that there exists an instance that can fall into the weighted minimum set cover problem.
\wsq

\section{Proposed Algorithm}   \label{sec:algorithm}
To efficiently solve the problem in this paper, we propose \textsc{ValidEOD}.
(As this algorithm outputs an arbitrary $\mathbf{E'}$ (if necessary), it is regarded as a heuristic solution for our problem under the constraint considered in Section \ref{sec:hardness}.)

\vsp
\noindent
\underline{\textbf{Main idea.}}
This algorithm improves the efficiency of the na\"ive algorithm by leveraging the fact that, even if a given eOD $\mathbf{E}: A \mapsto_{\leq} B$ does not hold, we need to compute an $\mathbf{E}' \supset \mathbf{E}$ from only pairs of tuples violating the eOD.
This needs a cheaper cost, because what we have to do is to check whether these tuples have missing values or not.
(If not, these tuples cannot be removed on any $\mathbf{E}' \supset \mathbf{E}$.)

We introduce an example with a swap that illustrates this main idea by using Table \ref{tab:eODSample}.

\begin{example} \label{example:main-idea}
Suppose that we want to validate $AB: A \mapsto_{\leq} B$.
From Theorem \ref{theorem:interchange}, it is sufficient to validate $AB: B \mapsto_{<} A$.
In Table \ref{tab:eODSample}, $B\mapsto_{<} A$ has a swap because of tuples $t_2$ and $t_3$, and this still exists on $\mathbf{r}^{AB}$.
Therefore, $AB: A \mapsto_{\leq} B$ is invalid.
We then want to compute a possible $\mathbf{E}' \supset \mathbf{E}$ such that the OD holds.
We need to add attributes with missing values into this embedding.
We add $D$ (the attribute in which $t_2$ has a missing value) to the embedding $AB$, and validate $ABD: B \mapsto_{<} A$.
As $\mathbf{r}^{ABD}$ does not contain $t_2$, the swap disappears.
We then see that $ABD: B \mapsto_{<} A$ is valid (and consequently $ABD: A \mapsto_{\leq} B$ is also valid) without enumerating a number of embeddings.
\end{example}

\begin{table}[!t]
    \centering
    \caption{Sample table for ValidEOD}
    \label{tab:eODSample}
    \begin{tabular}{cccccccc} \toprule
          ID    &  $A$  & $B$  & $C$ & $D$     & $F$     & $G$      & $H$       \\ \midrule
          $t_1$ &  4    & 1    & 1   &  8      & 20      &  10      & 1         \\ 
          $t_2$ &  6    & 2    & 3   & $\perp$ & 30      & $\perp$  & 2         \\ 
          $t_3$ &  5    & 3    & 5   &  10     & $\perp$ &  50      & 3         \\ 
          $t_4$ &  7    & 4    & 5   &  12     & 40      &  100     & $\perp$   \\ \bottomrule
    \end{tabular}
\end{table}

\noindent
\underline{\textbf{Overview.}}
\textsc{ValidEOD} has two phases.
It first validates a given eOD.
If it does not hold, then \textsc{ValidEOD} computes a new embedding (if it exists) for which the OD would hold.

\vsp
\noindent
\underline{\textbf{First phase: Validating eOD.}}
\textsc{ValidEOD} validates the given eOD in the following way.
First, it obtains a set $\mathbf{N}$ of attributes with missing values.
It then finds all swaps and merges for OD $B \mapsto_{<} A$, which is implemented by \textsc{FindErrors}.
This runs in linear time with respect to the number of rows in the relational table and returns sets $\mathbf{S}$ and $\mathbf{M}$ of swaps and merges, respectively. 
Then, \textsc{ValidEOD} checks whether all errors disappear under the given embedding $\mathbf{E}$ via \textsc{CheckForErrorDeletion}.
That is, it evaluates whether the found swaps and merges still exist on $\mathbf{r}^{\mathbf{E}}$. 
It is trivial that \textsc{CheckForErrorDeletion} needs a cost proportional to $|\mathbf{S}|$ and $|\mathbf{M}|$.
If no errors remain, \textsc{ValidEOD} returns valid.
Otherwise, it proceeds to the next phase.

\begin{algorithm}[!t]
    \caption{\textsc{UpdateEmbedding}}
    \label{algo:updateEmb}
    \DontPrintSemicolon
    \KwIn {$\mathbf{E}$: an embedding, $\mathbf{S}$: a set of tuples that cause swaps, $\mathbf{M}$: a set of tuples that cause merges, and $\mathbf{N}$: a set of attributes with missing values.}
    $\mathbf{E}' \leftarrow \mathbf{E}$\;
    \ForEach {$s \in \mathbf{S}$}
    {
        \ForEach {$A \in \mathbf{N}$}
        {
            $\mathbf{F} \leftarrow \mathbf{E}' + A$\;
            \If {$s$ is removed on $\mathbf{r}^{\mathbf{F}}$ }
            {
                $\mathbf{E}' \leftarrow \mathbf{F}$\;
                $\mathbf{S} \leftarrow \mathbf{S} - s$\;
                \textbf{break}
            }
        }
        \textbf{if} \textit{$s \in \mathbf{S}$} \textbf{then} \textbf{return} not valid
    }
    \ForEach {$m \in \mathbf{M} $}
    {
        \ForEach {$A \in \mathbf{N}$}
        {
            $\mathbf{F} \leftarrow \mathbf{E}' + A$\;
            \If {$m$ is removed on $\mathbf{r}^{\mathbf{F}}$}
            {
                $\mathbf{E}' \leftarrow \mathbf{F}$\;
                $\mathbf{M} \gets \mathbf{M} - m$\;
                \textbf{break}
            }
        }
        \If {$m \in \mathbf{M}$}
        {
            \textbf{return} not valid
        }
    }
    \textbf{return} valid under $\mathbf{E}'$
\end{algorithm}

\vsp
\noindent
\underline{\textbf{Second phase: New embedding computation.}}
Next, \textsc{ValidEOD} searches for an embedding $\mathbf{E}' \supset \mathbf{E}$ such that $\mathbf{E}': A \mapsto_{\leq} B$ holds, through \textsc{UpdateEmbedding}, which is described in Algorithm \ref{algo:updateEmb}.
As mentioned before, we need to care about whether pairs of tuples that have swaps or merges can be removed by adding attributes to a new embedding.
If such tuples have no missing values, they absolutely violate the given OD.
On the other hand, if the tuples have missing values, we may be able to remove the violations by updating the embedding.
Based on this idea, we incrementally update $\mathbf{E}$ so that swaps and merges are removed.
\textsc{UpdateEmbedding} specifically updates $\mathbf{E}$ as follows:
\begin{enumerate}
    \setlength{\leftskip}{-3.0mm}
    \item   For each swap $s \in \mathbf{S}$, \textsc{UpdateEmbedding} checks whether adding an attribute from $\mathbf{N}$ to the embedding removes $s$.
            If true, \textsc{UpdateEmbedding} adds this attribute to a new embedding $\mathbf{E}'$ (initialized by $\mathbf{E}$) and removes $s$ from $\mathbf{S}$.
            Otherwise, \textsc{UpdateEmbedding} tests the next attributes.
            If these tests still cannot remove $s$, it is guaranteed that $\mathbf{S} \neq \varnothing$.
            In this case, \textsc{UpdateEmbedding} returns not valid.
    \item   If \textsc{UpdateEmbedding} does not return not valid in the above step, it runs the same operations for each merge $m \in \mathbf{M}$.
    \item   If all swaps and merges are removed in the above steps, \textsc{UpdateEmbedding} returns valid with $\mathbf{E}'$.
\end{enumerate}

\noindent
\underline{\textbf{Space complexity}} of \textsc{ValidEOD} is trivially $O(|\mathbf{S}| + |\mathbf{M}| + |\mathbf{N}|)$.

\vsp
\noindent
\underline{\textbf{Time complexity.}}
The first phase needs $O(n + |\mathbf{S}| + |\mathbf{M}|)$ time, where $n$ is the number of rows in $\mathbf{r}^{\mathbf{E}}$.
The second phase needs $O(|\mathbf{N}|(|\mathbf{S}|+|\mathbf{M}|))$ time.
Thus, the time complexity of $\textsc{ValidEOD}$ is $O(n + (|\mathbf{N}|(|\mathbf{S}| + |\mathbf{M}|)))$.
That is, the time of \textsc{ValidEOD} is linear to the number of attributes, removing the \textit{factorial} cost held by the na\"ive algorithm.

\section{Experiment}   \label{sec:experiment}
This section reports our experimental results.
All experiments were conducted on a Ubuntu 20.04 LTS machine with 2.4GHz Intel Core i9-12900 and 64GB RAM.

\vsp
\noindent
\underline{\textbf{Datasets.}}
We used two real-world datasets, Adult (\url{https://archive.ics.uci.edu/}) and NCVoter (\url{https://www.ncsbe.gov/}).
The existing works \cite{jin2022efficient,langer2016efficient,wei2021algorithms} also used these datasets (but they removed tuples with missing values).
Adult consists of 32,000 tuples with 15 attributes and has 4,262 missing values, whereas NCVoter consists of 256,000 tuples with 19 attributes and has 796,496 missing values.

\vsp
\noindent
\underline{\textbf{Evaluated algorithms.}}
We compared our algorithm with the na\"ive algorithm.
Since this is the first work that deals with eODs, there are no other algorithms that can deal with eODs (i.e., we do not have existing competitors).
These algorithms were single-threaded, implemented in C++, and compiled by g++ 9.4.0 with -O3 flag.

\vsp
\noindent
\underline{\textbf{Parameters.}}
Given $\mathbf{X} \mapsto_{\leq} \mathbf{Y}$, $\mathbf{X}$ (resp. $\mathbf{Y}$) is called the left-hand (resp. the right-hand) side or LHS (resp. RHS).
To measure the performance of each algorithm, we varied the sizes of LHS and RHS.
The default sizes of LHS and RHS were one.
When we varied the size of LHS (resp. RHS), we fixed that of RHS (resp. LHS).
For each test, we randomly generated LHS and RHS from the attribute set and repeated this 10 times.

\begin{figure}[!t]
    \begin{center}
        \subfigure[Average running time vs. LHS size on Adult]{%
    		\includegraphics[width=0.47\linewidth]{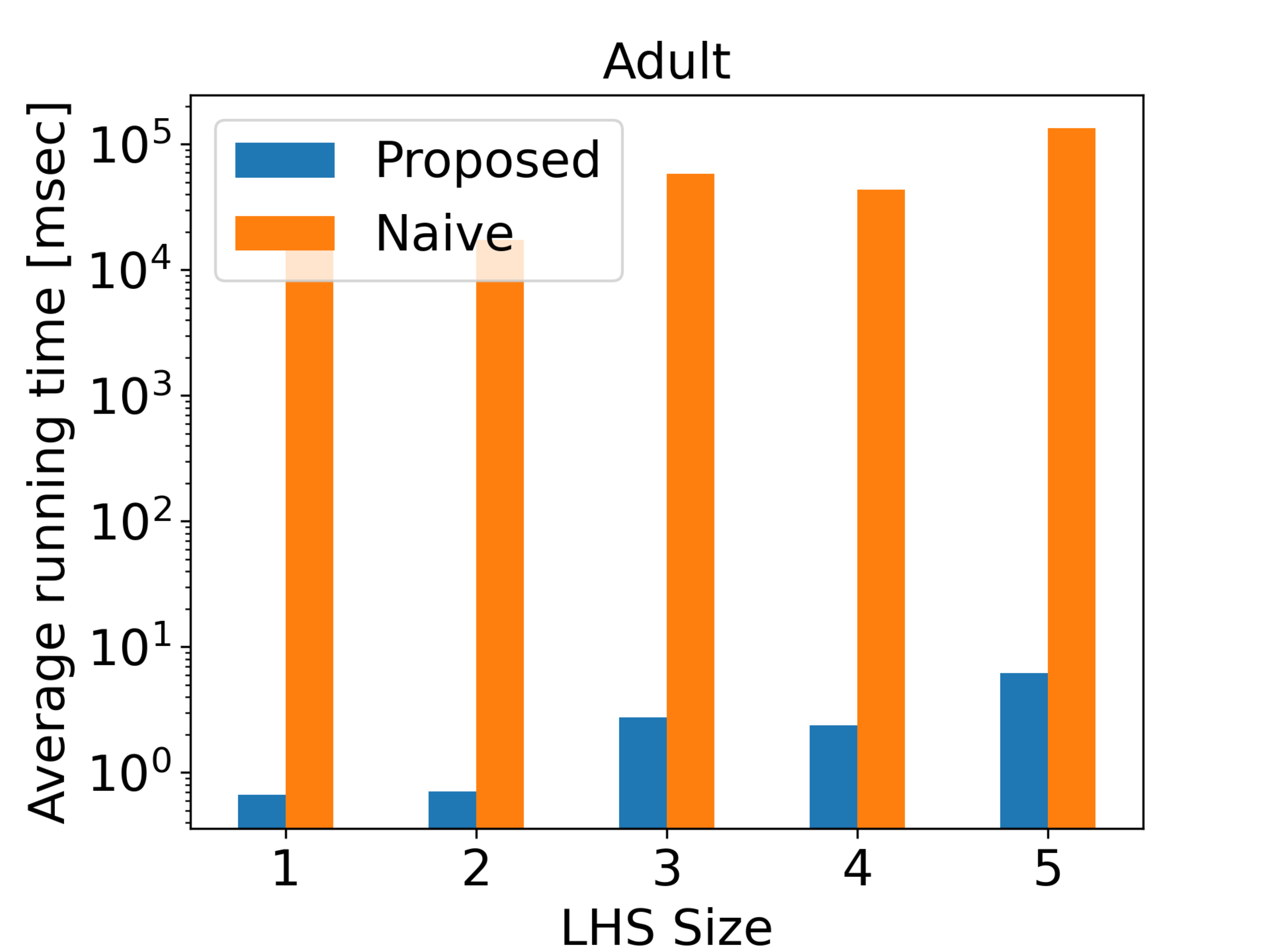}     \label{fig:AdultLHSAvg}}
        \subfigure[Average running time vs. RHS size on Adult]{%
    		\includegraphics[width=0.47\linewidth]{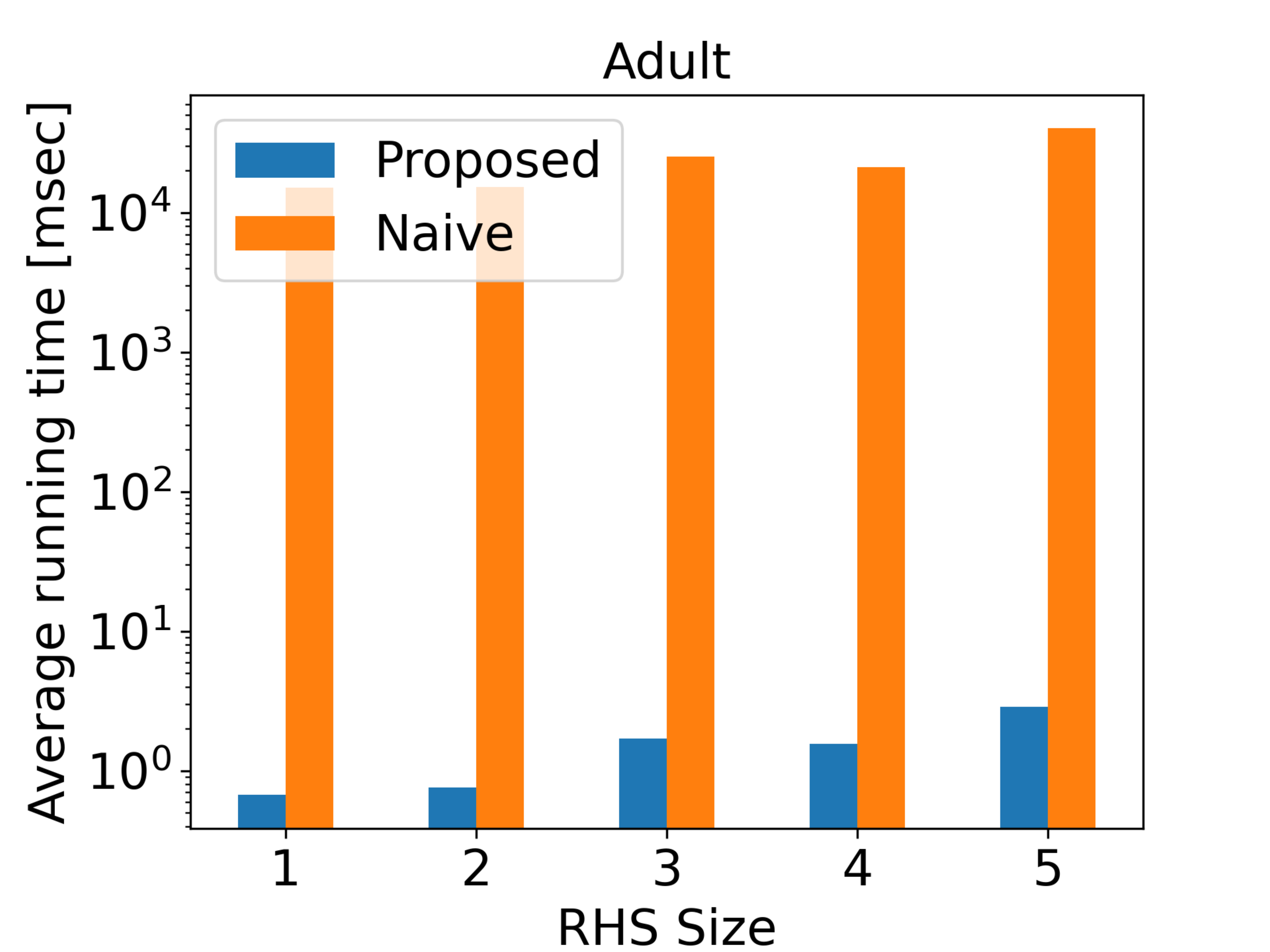}     \label{fig:AdultRHSAvg}}
        \subfigure[Average running time vs. LHS size on NCVoter]{%
    		\includegraphics[width=0.47\linewidth]{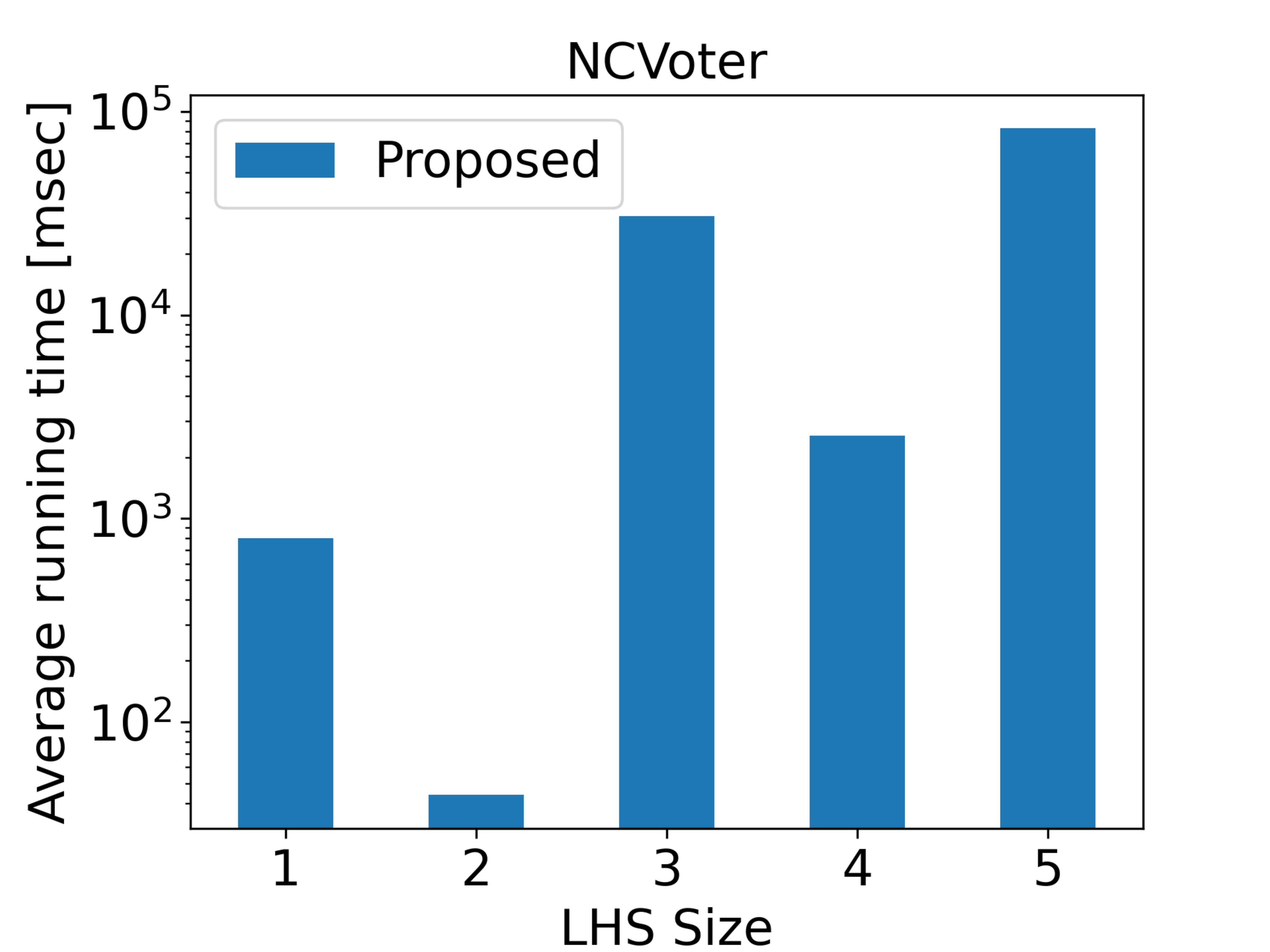}      \label{fig:NCVoterLHSAvg}}
        \subfigure[Average running time vs. RHS size on NCVoter]{%
    		\includegraphics[width=0.47\linewidth]{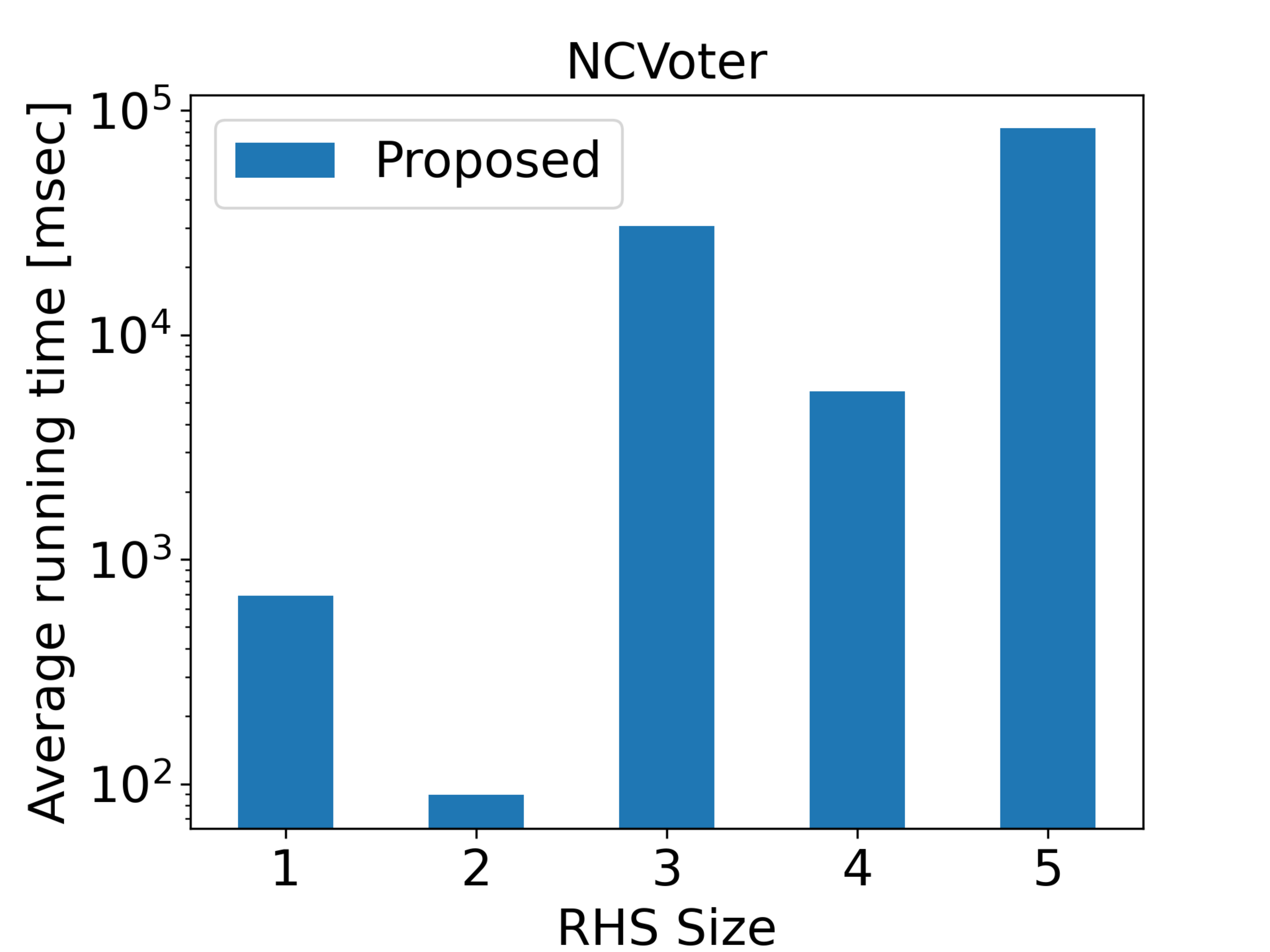}      \label{fig:NCVoterRHSAvg}}
    \caption{Running time vs. LHS and RHS sizes}
    \label{fig:Adult}
    \end{center}
\end{figure}

\vsp
\noindent
\underline{\textbf{Result.}}
Figure \ref{fig:Adult} shows how the na\"ive and our algorithms scale with respect to the LHS and RHS sizes.
The first observation is that the proposed algorithm is several orders of magnitude faster than the na\"ive algorithm.
Note that we omit the result of the na\"ive algorithm on NCVoter because it did not terminate on them within a few hours even for a single experiment (i.e., 10 iterations).
The datasets have more than 10 columns, and the factorial cost is huge.
Therefore, this result is reasonable.

We also observe that our algorithm is not affected by the LHS and RHS sizes.
Actually, this is expected from our theoretical analysis.
The time of our algorithm is dependent on the distributions of $\mathbf{S}$ and $\mathbf{M}$, not on the LHS and RHS sizes, see our time complexity analysis.

\begin{figure}[!t]
    \begin{center}
        \subfigure[Average $|\mathbf{S}|$ and $|\mathbf{M}|$ vs. LHS size on Adult]{%
    		\includegraphics[width=0.47\linewidth]{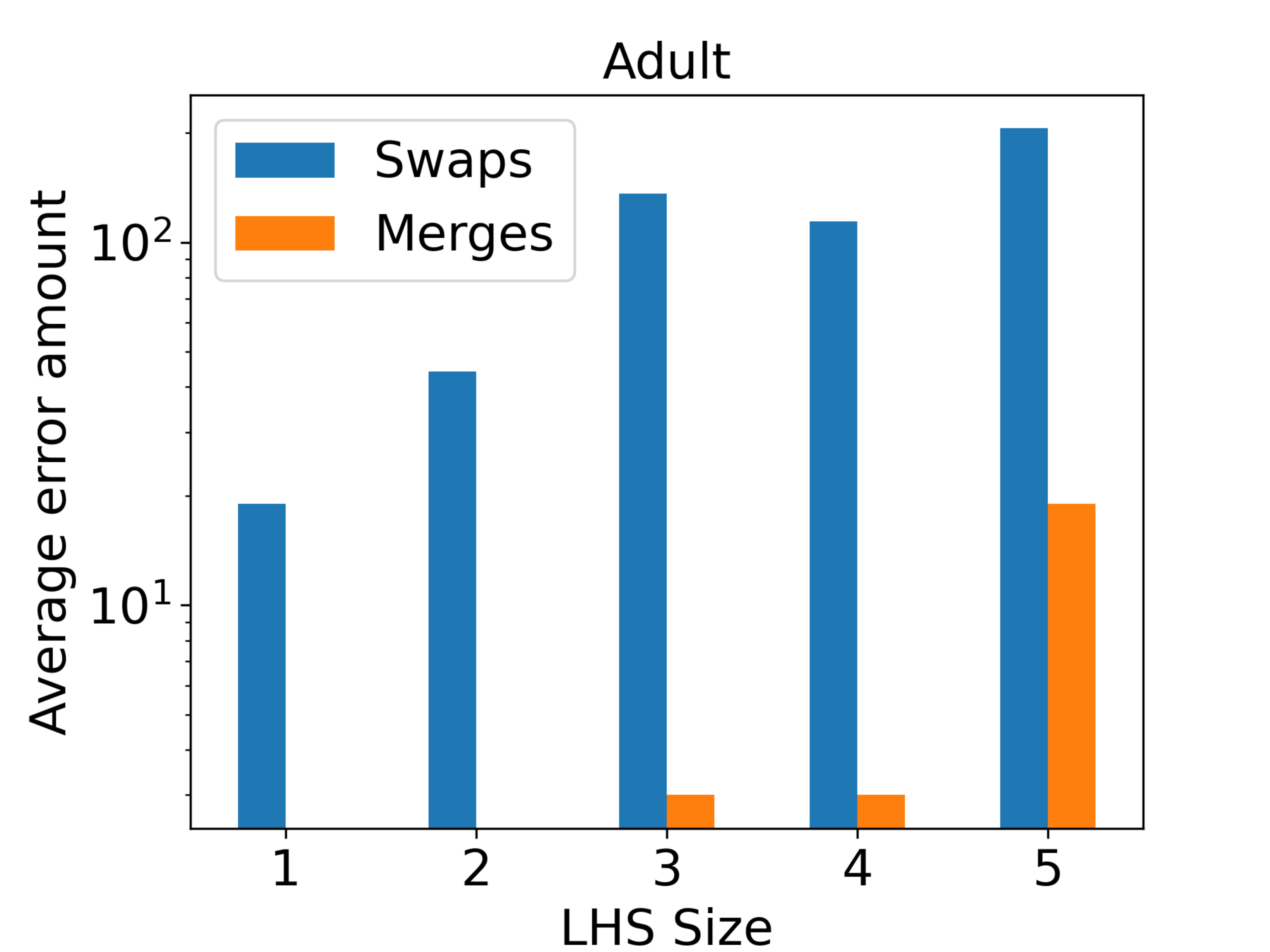}             \label{fig:AdultErrorsLHS}}
        \subfigure[Average $|\mathbf{S}|$ and $|\mathbf{M}|$ vs. RHS size on Adult]{%
    		\includegraphics[width=0.47\linewidth]{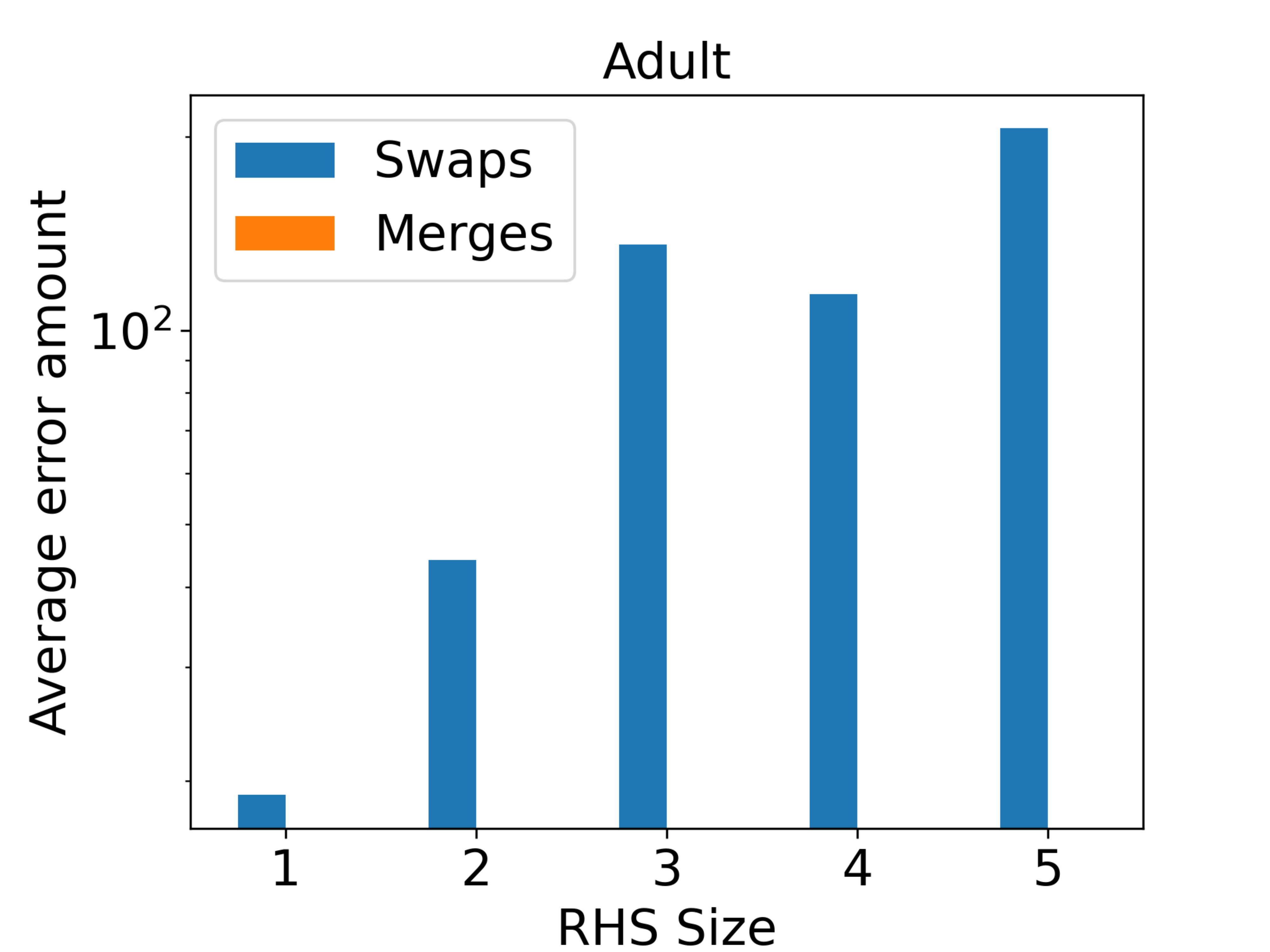}             \label{fig:AdultErrorsRHS}}
        \subfigure[Average $|\mathbf{S}|$ and $|\mathbf{M}|$ vs. LHS size on NCVoter]{%
    		\includegraphics[width=0.47\linewidth]{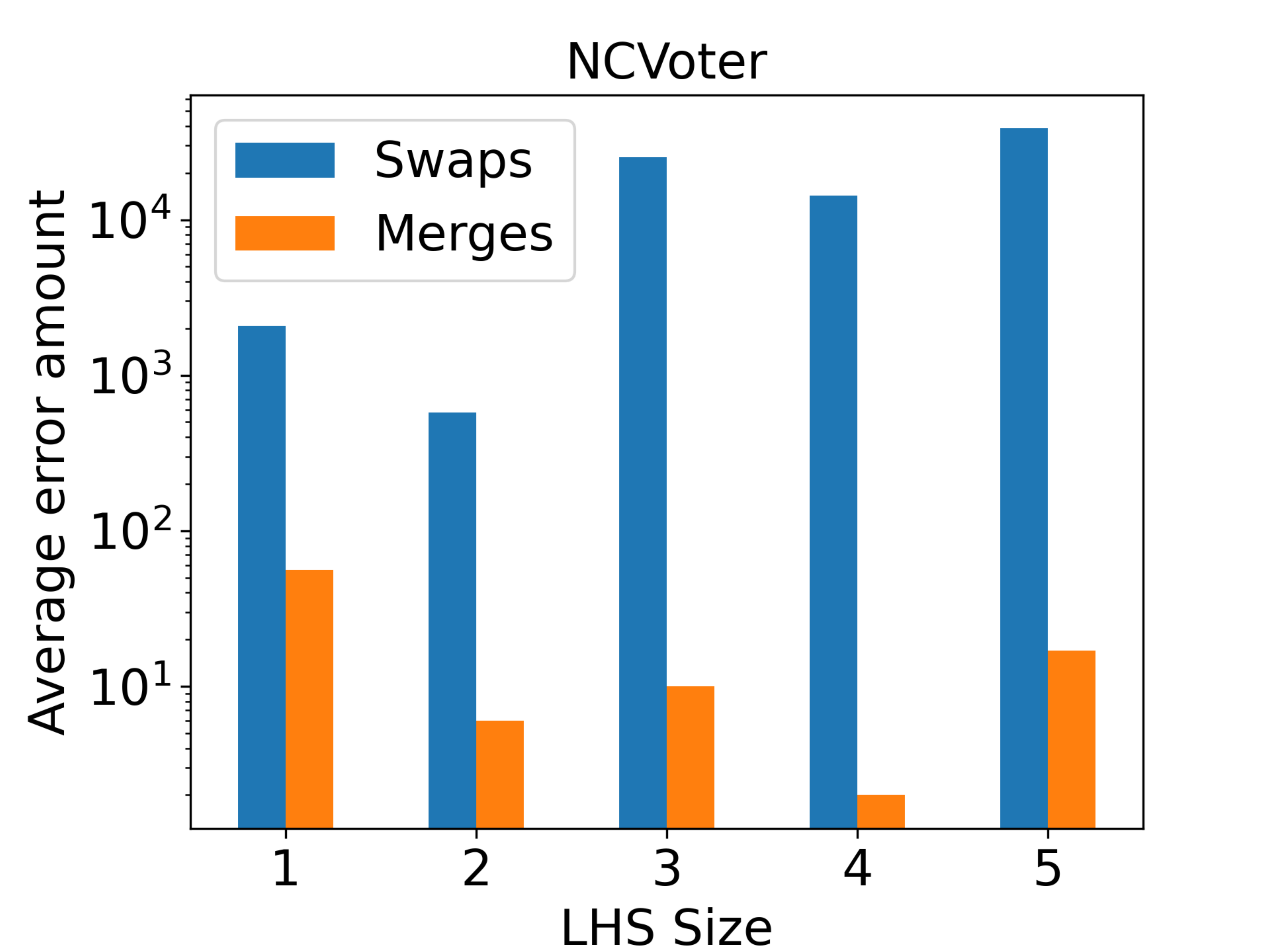}           \label{fig:NCVoterErrorsLHS}}
        \subfigure[Average $|\mathbf{S}|$ and $|\mathbf{M}|$ vs. RHS size on NCVoter]{%
    		\includegraphics[width=0.47\linewidth]{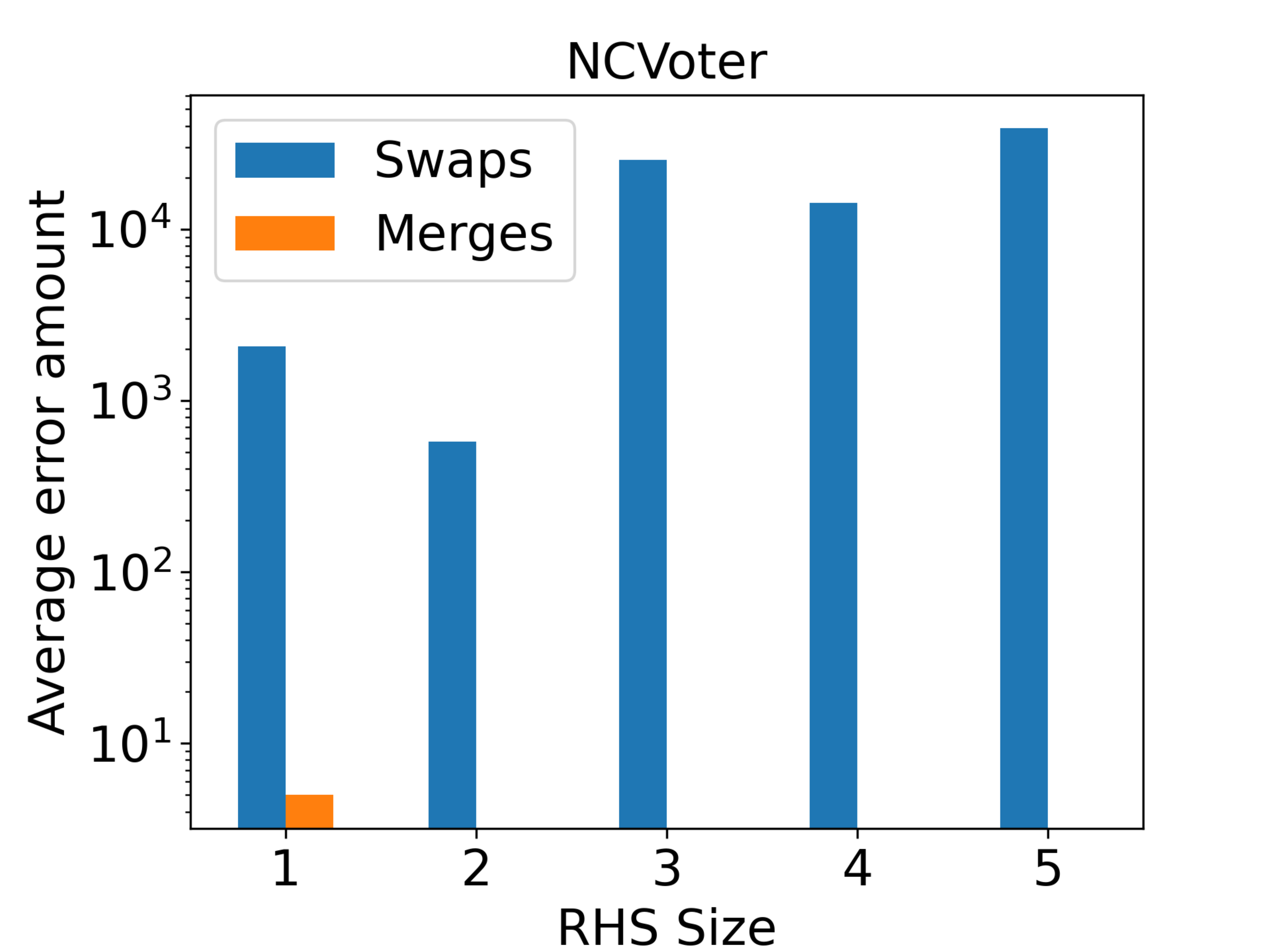}           \label{fig:NCVoterErrorsRHS}}
    \caption{Average $|\mathbf{S}|$ and $|\mathbf{M}|$ vs. LHS and RHS sizes}
    \label{fig:errors}
    \end{center}
\end{figure}

As stated above, our algorithm runs in time proportional to $|\mathbf{S}| + |\mathbf{M}|$.
Figure \ref{fig:errors} shows the average $\mathbf{S}$ and $\mathbf{M}$ on each dataset.
By comparing Figure \ref{fig:Adult} with Figure \ref{fig:errors}, it is clear that the running time of our algorithm follows the tendencies in Figure \ref{fig:errors}.
This result empirically validates our theoretical analysis in Section \ref{sec:algorithm}.

\section{Related Work}  \label{sec:related-work}

\subsection{Exact ODs}
Langer and Naumann presented ORDER \cite{langer2016efficient}, which finds all minimal ODs that hold in a given table.
ORDER itself is intentionally incomplete (as discussed in \cite{szlichta2016effective}).
Although it uses many pruning techniques that dramatically decrease the running time, it has the factorial worst-case time complexity with respect to the number of attributes.
Later, Schlichta et al. presented FASTOD \cite{szlichta2016effective}, an OD discovery algorithm with the exponential worst case time complexity with respect to the number of attributes and linear complexity with respect to the number of tuples.
They achieved this by mapping ODs to a set-based canonical representation. 

Exact ODs are useful for data with no errors (missing values).
However, there is a possibility that many potentially useful ODs exist but can never be found by this approach (whereas they may be found by ours or other approximate ODs).
Notice that ODs are a special case of eODs when the embedding is equal to all attributes in a dataset.

\subsection{Approximate ODs}
This topic (AODs) was defined in \cite{szlichta2016effective}.
Since ODs may not hold on datasets with errors, \cite{szlichta2016effective} considers the minimum number of tuples that must be removed from a given table for the OD to hold.
However, this problem is computationally expensive with respect to the numbers of tuples and attributes. 
Also, in \cite{jinA2021aproximate}, Jin et al. formalized the AOD discovery problem and developed efficient algorithms for AOD discovery with an error measure optimization.

Although AODs find ODs that hold on dirty data, they do not consider data with missing values.
Hence, AODs cannot provide the completeness requirement that helps data cleaning and data schema design (whereas eODs can do this).

\subsection{Embedded Functional Fependencies}
Wei and Link introduced the concept of Embedded Functional Dependencies (eFDs) \cite{wei2019embedded}.
These are used to establish a robust schema design framework independent of the interpretation of missing values.
In \cite{wei2021algorithms}, they present row-efficient, column-efficient, and hybrid approaches for discovering eFDs.

Although eFDs provide valuable data completeness and data integrity requirements, they do not consider integrity requirements with respect to order.
Since ODs subsume FDs, eODs also subsume eFDs.

\section{Conclusion}    \label{sec:conclusion}
This paper proposed a new concept of embedded order dependencies to deal with order dependencies with missing values and to satisfy the integrity and completeness requirements.
A na\"ive algorithm incurs a factorial cost with regard to the number of attributes, so it does not scale well to large relational databases.
Motivated by this, we presented an efficient algorithm that checks whether an OD holds on a given embedding and returns (if possible) an embedding on which it holds.
We conducted experiments on real-world datasets, and the results demonstrate the efficiency of our eOD validation algorithm.

\begin{acks}
This research is partially supported by AIP Acceleration Research JPMJCR23U2 and JST CREST JPMJCR21F2. 
\end{acks}

\bibliographystyle{ACM-Reference-Format}
\bibliography{bibtex}

\end{document}